\documentclass{elsart}
\usepackage{graphicx}

\def\textbf#1{{\bf #1}}
\def\textit#1{{\it #1}}

\begin{document}

\begin{frontmatter}

\title{ A mechanism leading from bubbles to crashes: the case of Japan's land market}
\author[Tokyo1,Tokyo2]{Taisei Kaizoji\thanksref{contact}}
\author[Tokyo2]{Michiyo Kaizoji}
\address[Tokyo1]{%
  Division of Social Sciences,
  International Christian University, Osawa, Mitaka, Tokyo 181-8585,
  Japan}
\address[Tokyo2]{%
  Econophysics Laboratory, 5-9-7-B Higashi-cho, Koganei-shi, Tokyo, 184-0011, Japan.}
\thanks[contact]{%
Corresponding author. E-mail: kaizoji@icu.ac.jp}

\begin{abstract}
In this study we investigate quantitatively statistical properties of an ensemble of {\it land prices} in Japan in the period from 1981 to 2002, corresponding to a period of bubbles and crashes. We found that the tail of the complementary cumulative distribution function of the ensemble of land prices in the high price range is well described by a power-law distribution, $ P(S>x) \sim x^{-\alpha} $, and furthermore that as the power-law exponents $ \alpha $ approached unity, bubbles collapsed.  
\end{abstract}
\begin{keyword}
econophysics \sep power law \sep 
ensemble distribution
\PACS 89.90.+n \sep 05.40.-a
\end{keyword}

\end{frontmatter}

\section{Introduction}
The burst of the Japanese real estate bubbles in the early 1990s was the bust following the greatest boom of the late twentieth century. The average land price increased 550 percent between 1981 and 1991. See Figure 1(a). By 1991, the total value of all Japanese property was estimated at nearly 20 trillion dollars, which equaled more than 20 percent of the world's wealth. Theoretically, the Japanese could have bought all the property in America by selling off metropolitan Tokyo \cite{1}. This anomalous appreciation of land price was created by an irrational exuberance of land speculation that was fueled by a myth that the land prices could never go down in Japan. In 1991, however, the bottom fell out of the castle-in-air market. From 1991 to 2002, the average land price continued . By 2002 the average land price fell by about 70 percent compared with the level in 1991. Figure 1(b) shows the movement of the mean value of the relative change in land prices $ R $. \par
Why did the real estate bubble collapse crash? Here we analyze the ensemble distribution of land price in Japan and offer an explanation of a mechanism leading from bubbles to crashes. 
In our previous work \cite{2} we investigated a database of the {\it assessed value of land} that is made public once a year by the Ministry of Land, Infrastructure and Transport Government of Japan (http://www.mlit.go.jp/ksj/) and found that the complementary cumulative distributions of ensembles of the land prices fits to a very good degree of approximation a power-law distribution\footnote{Yamano \cite{3} found that complementary cumulative distribution of the absolute value of year-on-year percentage changes of land prices has been motivated by the generalized thermostatistics.}. Andersson, et. al. \cite{4} proposed a network model of urban economy that assumed that the market dynamics that generate land values can be represented as a growing scale-free network. We may, therefore, reasonably conclude that the distribution of land prices is accurately described by a power-law distribution. We have attempted in this paper to extend the observation into the 22-years period between 1981 and 2002, to explain a mechanism leading bubble to crashes. 

\section{Power-law for ensemble of land prices}
In Figure 1(c), we plot the cumulative probability distribution of ensembles of land prices (S) in the four years 1985, 1987, 1991, and 1998, on the log-log scale. The data for each of the 22 annual intervals from 1981 to 2002 are for land prices over 10,000 points. We found that the tail of the complementary cumulative distribution function of the ensemble of the land prices in the high price range is well described by a power-law distribution, $ P(S>x) \sim x^{-\alpha} $. We use ordinary least squares (OLS) regression in log-log coordinates in order to determine the exponent $\alpha$ for the data of land prices for each of 22 years from the period 1981 - 2002. Overall, the power-law distribution of land prices is very strong. Figure 1(d) indicates the movement of the exponent $ \alpha$. The exponent $\alpha$ continued to decrease toward unity between 1981 and 1987, and during the period of the peak of bubbles from 1987 to 1992 the exponent $\alpha$ hovered around unity. In 1991 the bubbles crashed, the trend has reversed, and the exponent $\alpha$ continued to increase between 1992 and 2001. This finding suggests that the threshold value of the exponent $\alpha$ that cause bubbles to burst is unity. We interpret this findings as follows. The mean of the power-law distribution is equal to $ 1/(\alpha - 1) $. Therefore, as the power-law exponent $ \alpha $ approaches unity, the mean diverges and the bubbles collapse. 

From a practical viewpoint let us then consider these empirical findings. It is well known that the exponent $\alpha$ of the power-law distribution can be considered as a measurement of wealth inequality in real estate holdings. The Gini coefficient $(G)$ \cite{5}, known as the index for wealth concentration, can be written as $G=1/(2 \alpha - 1) $. Theoretically, the Gini coefficient ranges from zero, when all land areas are equal in price, to unity, when one land area has the highest price and the rest none. If the exponent $ \alpha $ is close to unity, it means the Gini coefficient will be close to unity. Therefore, the bubble is not defined as the generally extraordinary rise of land prices, but rather as the extraordinary enlargement of the inequality of land prices. Figure 1(e) shows the movement of the Gini coefficient. The inequality of the distribution of land prices measured by the Gini coefficient increased drastically from 0.58 in 1981 to 0.77 in 1988, and particularly during the period of 1987 - 1992 the Gini coefficient is bigger than 70 percent, and inequality reached the breaking point. The real estate bubbles caused intolerable inequality for Japanese society. Accordingly, increasing the wealth inequality acted as a trigger to cause the collapse of the real estate bubble. 

\section{Concluding remarks} 

Many empirical findings of asset time series, which have been proposed recently, have suggested that asset markets including land markets are complex systems in which a large number of agents participate in trade and have interplay with each other. Assuming asset markets are complex systems, crashes in asset markets will be considered as catastrophes in complex systems. Although it is extremely important to elucidate the nature of any catastrophe in complex systems, little is known about the mechanism of such catastrophes. In this study, we investigated land markets quantitatively, and showed causes of market crashes empirically. The next step is to model this finding. The theoretical study will be left for future work. 

\section{Acknowledgement}

Financial support by the Japan Society for the Promotion of Science under the Grant-in-Aid, No. 06632 is gratefully acknowledged. All remaining errors, of course, are ours.


\begin{figure}
\begin{center}
  \includegraphics[height=15cm,width=14cm]{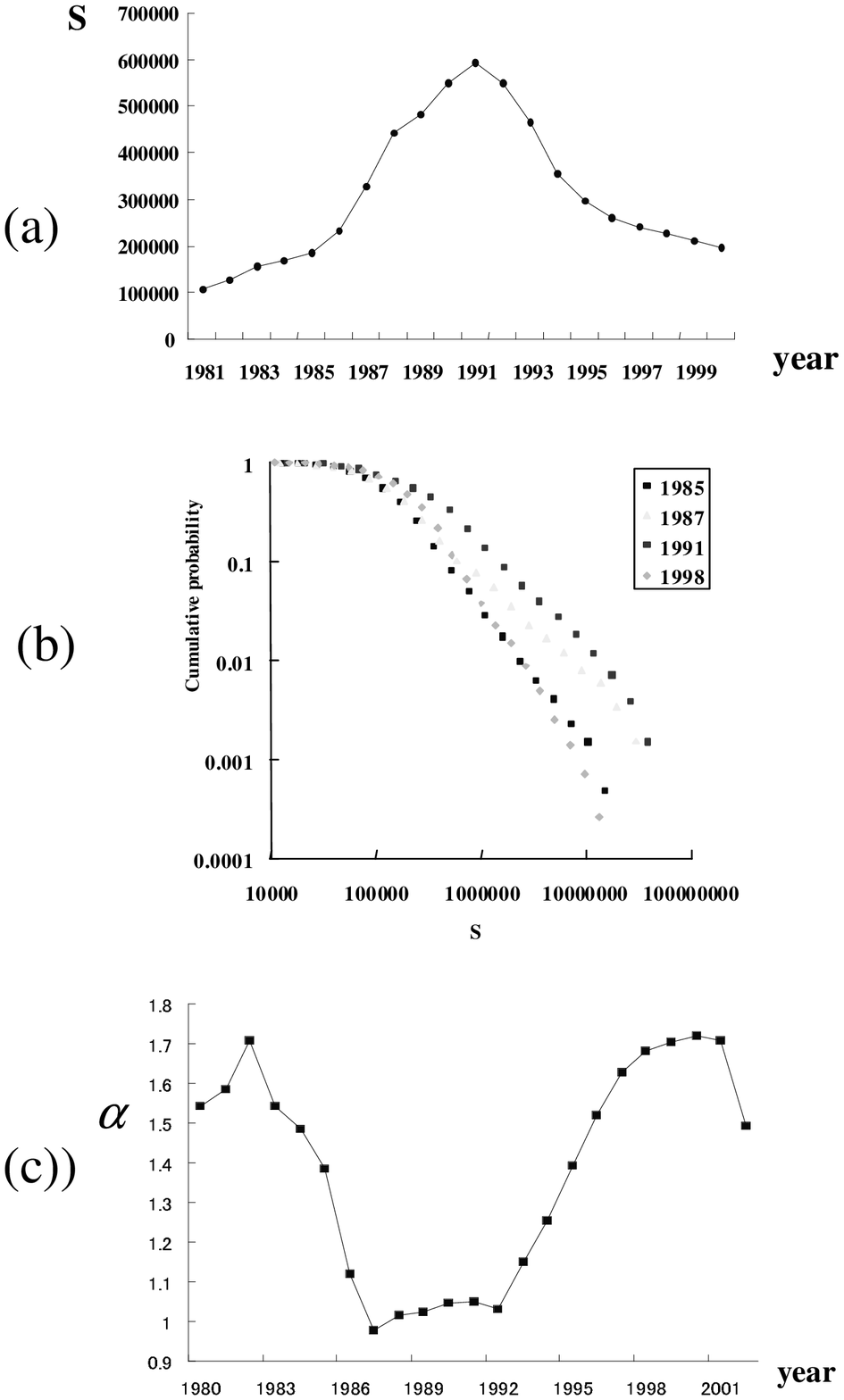}
\end{center}
\caption{(a) Log-log plot of the cumulative distribution of ensembles of the land prices for each of the four years 1985, 1987, 1991, and 1998.The movement of the mean values of land prices. (b) The movement of power-law exponent $ \alpha $ of land prices' ensemble distribution. (c) The movement of coefficient of variation (CV) of land prices' ensemble distribution.  (d) Log-log plot of the cumulative distribution of ensembles of the stock prices on January 4, 2000. (e) The movement of power-law exponent $ \alpha $ of stock prices' ensemble distribution. (f) The movement of coefficient of variation (CV) on stock prices' ensemble distribution.}
\label{fig1}
\end{figure}
\end{document}